\documentclass[fleqn,12pt,twoside]{article} 
\usepackage{espcrc1} 
\usepackage{epsfig} 
 
\usepackage{graphicx} 
\usepackage[figuresright]{rotating} 
 
\newcommand{\AmS}{{\protect\the\textfont2 
  A\kern-.1667em\lower.5ex\hbox{M}\kern-.125emS}} 
 
\title{Mapping Deconfinement with 
a Compact Star Phase Diagram} 
 
\author{H. Grigorian\address[rostock]{Rostock University, Germany}, 
\address[yerevan]{Yerevan State University, Armenia}, 
    \thanks{Supported by DFG Grant No: 436 ARM 17/5/01}, 
        D. Blaschke\addressmark[rostock],\address[dubna]{JINR Dubna}, 
        G. Poghosyan\addressmark[yerevan], 
    \thanks{Supported by DAAD Grant No: A/01/08510 for postdoctoral 
    studies at Rostock University}} 
\begin{document} 
 
\maketitle 
 
\begin{abstract} 
We have found correlations between properties of the equation of 
state for stellar matter with a phase transition at supernuclear 
densities and two characteristic features of a {\it phase diagram} for 
rotating compact stars in the angular velocity - 
baryon number plane: 1) the critical dividing line between mono- 
and two-phase star configurations and 2) the maximum mass line. 
The second line corresponds to the minimum mass function for black 
hole candidates whereas the first one is observable by a population 
statistics, e.g. for Z-sources in low-mass X-ray binaries. 
The observation of a population gap in the mass distribution for 
the latter is suggested as an astrophysical verification of the existence 
of a first order phase transition in QCD at high densities such as 
the deconfinement. 
\end{abstract} 
 
\section{Introduction} 
 
At present, the existence of exotic phases of matter at 
high densities is under experimental investigation in ultrarelativistic 
heavy-ion collisions \cite{qm01} the most prominent being the 
deconfined phase of QCD \cite{bkr}. 
While the diagnostics of a phase transition in experiments with heavy-ion 
beams faces the problems of strong nonequilibrium and finite size, 
the dense matter in a compact star forms a macroscopic system in 
thermal and chemical equilibrium for which effects signalling a 
phase transition shall be most pronounced. 
 
We introduce a classification of rotating compact 
stars in the plane of their angular frequency $\Omega$ and mass 
(baryon number $N$) which we will call {\it phase diagram}. 
In this diagram, configurations with high density matter cores are separated 
from conventional ones by a critical phase transition line. 
The position and the form of these lines are sensitive to changes 
in the equation of state of stellar matter \cite{nsi}. 
Since the phase diagram of rotating compact objects seems to be a 
more general approach for investigations of phase transition 
effects in the interior of the star we assume that the 
deconfinement transition could be a particular case besides of 
other possibilities for phase transitions like pion or kaon 
condensation as discussed, e.g. in \cite{migdal,reddy}. 
Therefore, our aim is to suggest it as a heuristic tool for obtaining 
constraints on the EoS at high densities from the rotational 
behaviour of compact stars. 
 
\section{EoS and compact star phase diagram} 
 
\begin{figure}[htb] 
\begin{minipage}[t]{80mm} 
\epsfig{figure=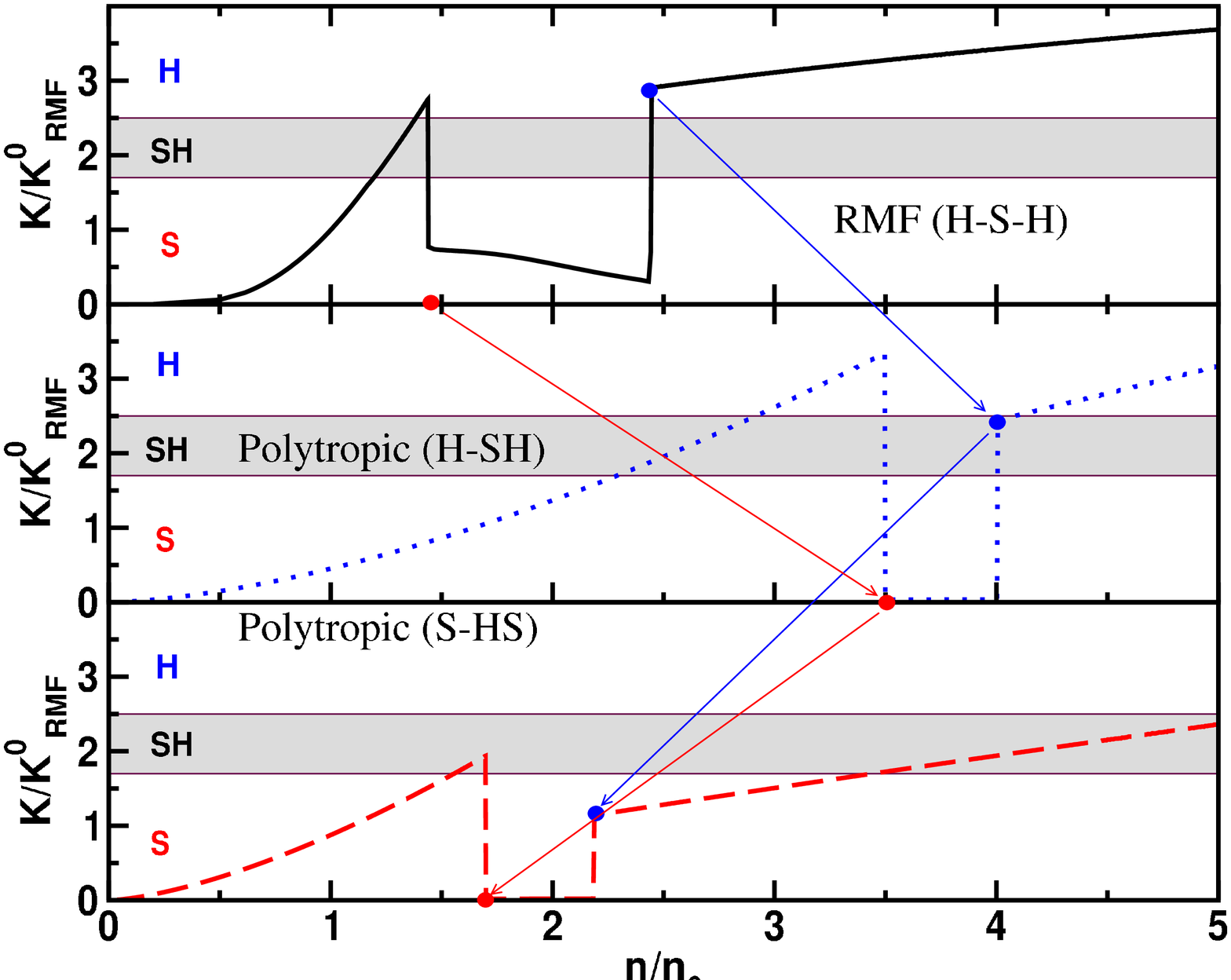,width=88mm} 
\caption{Incompressibilities for RMF and polytropic EoS models 
with a phase transition, see text and \cite{prl}.} 
\label{EoSfig} 
\end{minipage} 
\hspace{\fill} 
\begin{minipage}[t]{75mm} 
\epsfig{figure=phd.eps,width=77mm} 
\caption{Phase diagrams for rotating star configurations for 
the EoS of Fig. \ref{EoSfig}.} 
\label{phdiags} 
\end{minipage} 
\end{figure}

Our focus is on the elucidation of qualitative features of 
signals from the high density phase transition in the pulsar timing, 
therefore we use a generic form of an equation of state (EoS) with such a 
transition. 
We use the polytropic type  equation of state \cite{physrep} 
for different values of the incompressibility \cite{gbook} 
$K_{L,H}(n)=9~dP/dn$ at the saturation density, see Ref. \cite{prl}. 
The phase transition between the lower and higher density phases is made 
by the Maxwell construction \cite{hh} and compared to a relativistic mean 
field model consisting of a linear Walecka plus dynamical quark 
model EoS \cite{bbkr,bt00} with a Gibbs construction \cite{glen,cgpb}, 
see Fig.\ref{EoSfig}. 
 
In Fig. \ref{phdiags} we display the phase diagrams for the rotating star 
configurations, which correspond to the three model EoS of Fig. \ref{EoSfig}. 
These phase diagrams have four regions: 
(i) the region above the maximum frequency $\Omega >\Omega_{\rm K}(N)$ 
where no stationary rotating configurations are found, (ii) the region of 
black holes $N > N_{\rm max}(\Omega)$, and the region of stable compact 
stars which is subdivided by the critical line $N_{\rm crit}(\Omega)$ into 
(iii) the region of hybrid stars for $N > N_{\rm crit}(\Omega)$ where 
configurations contain a core with a second, high density phase and 
(iv) the region of mono-phase stars 
without such a core. 
 
From the comparison of the regional structure of these three different phase 
diagrams in Fig. \ref{phdiags} with the corresponding EoS 
in Fig. \ref{EoSfig} we arrive at the main result of this paper that 
there are the following correlations between the topology 
of the lines $N_{\rm max}(\Omega)$ and $N_{\rm crit}(\Omega)$ and the 
properties of two-phase EoS: 
\begin{itemize} 
\item[-] 
The hardness of the high density EoS determines the maximum mass of 
the star, which is given by the line $N_{\rm max}(\Omega)$. 
Therefore  $N_{\rm max}(0)$ is proportional to the parameter $K_H(n_H)$, 
where $n_H$ is the density of the transition to high density phase. 
\item[-] 
The onset of the phase transition line $N_{\rm crit}(0)$ depends on 
the density $n_H$ and $K_L(n_L)$ where $n_L$ is the density of the transition 
to the low density phase. 
\item[-] 
The curvature of the lines $N_{\rm max}(\Omega)$ and 
$N_{\rm crit}(\Omega)$ is proportional to the compressibility of the high and 
low density phases, respectively. 
\end{itemize} 
 
Therefore, a verification of the existence of the critical lines 
$N_{\rm crit}(\Omega)$ and $N_{\rm max}(\Omega)$ by observation of 
the rotational behavior of compact objects would 
constrain the parameters of the EoS for neutron star matter. 
Moreover, we have investigated different trajectories of rotating 
compact star evolution in the phase diagram in order to identify 
scenarios, which result in signatures of the deconfinement phase 
transition. 
To  prove that the appearance or disappearance of a high density phase 
during the rotational evolution of the star could entail 
observational consequences for the 
angular velocity we consider three main representatives different 
classes of trajectories 
which could cross the critical line 
on phase diagram. 
These classes of tracks are: 
(a) spindown of isolated (non-accreting, $\dot N=0$) 
pulsars due to magnetic dipole radiation \cite{frido,cgpb}, 
(b) spin up in accreting systems with weak magnetic field 
\cite{fridonew,accmag} ($N \simeq{\rm const}$, vertical tracks) and 
(c) accretion either with strong magnetic field \cite{accmag} 
or for accreting binaries emitting gravitational 
waves \cite{bildsten}, for which $\Omega \simeq {\rm const}$ 
(horizontal tracks). 
In the case of (a) the spindown $K= K_{\rm rad}$ or (b) spinup  $K= K_{\rm 
acc}$ evolutions (in both cases $K_{\rm int}<< K_{\rm ext}$) 
the  objects can undergo  a phase transition if the baryon number 
lies within the interval 
$N_{\rm crit}(\Omega = 0)< N < N_{\rm c}$, where $N_{\rm c}$ is the 
end piont of the critical line $N_{\rm crit}(\Omega)$. 
If the core of compact star is soft enough (as in case (SH-S)) 
the critical line $N_{\rm crit}(\Omega)$ crosses $N_{\rm 
max}(\Omega)$ at some $\Omega_c(N_c)< \Omega_K(N_c)$. 
This means that massive mono-phase configurations with total baryon number 
$N>N_c$ rotating with angular velocities in the interval 
$\Omega_c<\Omega < \Omega_K$ 
should encounter a transition to a black hole during the spin down evolution. 
In those cases when the core EoS is harder (H-S-H and SH-H), the region of hybrid 
stars is a band which separates mono-phase configurations from black holes, 
see upper two panels of Fig. \ref{phdiags}. 
As it has been shown in \cite{cgpb} for the vertical tracks (a) and 
(b) in the phase 
diagram, the braking index $n(\Omega)$ changes its value from $n(\Omega)>3$ 
in the region (iii) to $n(\Omega)<3$ in (iv). 
This is the braking index signal for a deconfinement transition 
introduced in Ref. \cite{frido}. 
The third evolutionary track is accretion with strong magnetic fields 
\cite{accmag} and/or gravitational wave emission (horizontal tracks) 
\cite{bildsten}.For this case the $\dot \Omega$ first 
decreases as long as the moment of 
inertia monotonously increases with $N$. When passing the critical 
line $N_{\rm crit}(\Omega)$ for the phase transition, the 
moment of inertia starts decreasing and the internal torque term 
$K_{\rm int}$ changes sign. This leads to a narrow dip for $\dot 
\Omega (N)$ in the vicinity of this line. As a result, the phase 
diagram gets overpopulated for $N \stackrel{<}{\sim} N_{\rm 
crit}(\Omega)$ and depopulated for $N \stackrel{>}{\sim} N_{\rm 
crit}(\Omega)$ up to the second maximum of $I(N, \Omega)$ close to 
the black-hole line $N_{\rm max}(\Omega)$. 
A {\it population gap} in the phase diagram of compact stars 
is appears as a detectable indicator for hybrid star configurations. 
 
\section{Conclusion} 
 
We have shown that population clustering of compact stars at the phase 
transition line could be a signal for the 
occurrence of stars with high density matter cores and a measure for 
obtaining constraints on the EoS at high densities. 
In the case without a phase transition, the 
moment of inertia could at best saturate before the transition to 
the black hole region and consequently $\dot \Omega$ would also 
saturate. This would entail a smooth population of the phase 
diagram without a pronounced structure \cite{ssacc}. 
The so called Z sources of QPOs in LMXBs are suggested as objects 
which should predominantly populate the region of the suspected 
phase border between hadronic stars and quark core stars in phase 
diagram \cite{cgpb}. 
The existence of a population gap between the critical line and 
the black hole limit on the phase diagram is suggested as a signal 
for a phase transition at supernuclear densities in accreting 
compact stars \cite{accmag}. The absence of a such gap does not exclude the 
existence of deconfined quark matter in LMXBs as we have shown in a 
study of accreting strange stars \cite{ssacc}.

\end{document}